	       \documentclass[multphys,vecphys]{svmult}
	\def\deg      {\ifmmode{^{\circ}}\else{$^{\circ}$}\fi}
\def\decdeg {\rlap . {}^\circ}     
			\def\z {\phantom{$0$}}
		       \def\zz {\phantom{$00$}}

\usepackage{makeidx}     
\usepackage{graphicx}    
\usepackage{multicol}    

\makeindex             

\begin{document}

\title*{Perspective from a Younger Generation --\\
The Astro-Spectroscopy of Gisbert Winnewisser}
\titlerunning{The Astro-Spectroscopy of Gisbert Winnewisser}
\author{Karl M. Menten}
\authorrunning{The Astro-Spectroscopy of Gisbert Winnewisser}
\institute{Max-Planck-Institut f\"ur Radioastronomie, Auf dem H\"ugel 69,
D-53121 Bonn, Germany
\texttt{kmenten@mpifr-bonn.mpg.de}}
%
%
\maketitle

\vspace{-6.5cm}
\noindent
\leftline{To appear in the }
\leftline{\tt \bf Proceedings of the 4th Cologne-Bonn-Zermatt-Symposium}
\leftline{\it ``The Dense Interstellar Medium in Galaxies''}
\leftline{eds. S. Pfalzner, C. Kramer, C. Straubmeier, \&\ A. Heithausen
(Springer: Berlin)} 
\vspace{-.2cm}
\hrule
\vspace{4.6cm}

Gisbert Winnewisser's astronomical career was practically coextensive with 
the whole development of molecular radio astronomy. Here I would like to
pick out a few of his many contributions, which I, personally, find 
particularly interesting and put them in the context of newer
results.
\section{Beginnings -- When Molecular Radio Astronomy was still 
\textit{Radio} Astronomy}
\label{sec:1}
\subsection{Introduction}
The late 1960s and early 70s were an era of great excitement
in all branches of radio astronomy. In spectral line research, 
12 years after the detection of the first line, 
the 21 cm HI hyperfine structure (hfs) transition  (Ewen \&\ Purcell 1951), finally
the first molecule was found, the hydroxyl (OH)
radical (Weinreb et al. 1963). 

The OH $\Lambda$-doublet transitions, as well as many other lines at m-, cm- and(!) mm-wavelengths
from a variety of molecules,  
had been identified as possible astronomical targets in a visionary article by
Charles Townes (1957)\footnote{see also Shklovsky 1952.}. 

Shortly after the detection of OH, 
Hoglund \&\ Mezger (1965) discovered the first radio recombination line,
H109$\alpha$ at 5009 MHz. \textit{Molecular}
radio astronomy started in earnest in  
1968/9, with the discoveries of ammonia (NH$_3$; Cheung et al. 1968) and
water vapor (H$_2$O;
Cheung et al. 1969) by a group around Townes\footnote{Both
lines were on Townes' (1957) list!}, as well as the 
first complex organic 
species,
formaldehyde (H$_2$CO; Snyder et al. 1969), breaking loose
a real gold rush that has yet to stop. 
Even today, approximately 
2 -- 4 ``new'' interstellar or circumstellar\footnote{Circumstellar
molecules exist in the envelopes of evolved giant stars. Many 
species are detected there that are also found in interstellar clouds, whereas
some others, have so far \textit{only} been detected in 
a circumstellar environment.} molecules are
found each year, totaling 126 species (not counting 
isotopomers)\footnote{See http://www.ph1.uni-koeln.de/vorhersagen/molecules/main\_molecules.html}.

The early history of ``molecule-hunting'', interesting us here, is well
documented by the review by Snyder (1972; ``Molecules in Space''), 
which gives a very useful introduction to molecular
radio astronomy, that is still very much readable today. In the same volume, the review following Snyder's
is by Winnewisser, Winnewisser, \&\ Winnewisser and presents an equally excellent  
overview of the laboratory side of ``Millimetre Wave Spectroscopy''.

\begin{table}
\begin{center}
\caption{The First Interstellar ``Radio'' Molecules}
\label{tab:1}       
\scriptsize{
\begin{tabular}{lllll}
\hline\noalign{\smallskip}
Name & Species & Transition & Frequency (MHz) & Reference  \\
\noalign{\smallskip}\hline\noalign{\smallskip}
Hydroxyl      &OH                &$^2\Pi_{3\over2}, J={3\over2}$  & \z1665.4018(2)$^{\rm a}$  & Weinreb et al. 1963\\
Ammonia       &NH$_3$            & $(1,1)$                      &  23694.4955(1)$^{\rm b}$    & Cheung et al. 1968\\
Formaldehyde  &H$_2$CO           & $1_{11}-1_{10}$              &  \z4829.660(1)$^{\rm b}$    & Snyder et al. 1969 \\
Water         &H$_2$O            & $6_{16}-5_{23}$              &   22235.08(2)$^{\rm b}$    & Cheung et al. 1969\\
Methyl alcohol&CH$_3$OH          & $1^+-1^-$                    &   \zz834.267(2)     & Ball et al. 1970\\
Formic acid   &HCOOH             & $1_{11}-1_{10}$              &   \z1638.804(1)     & Zuckerman et al. 1971\\ 
              &                  & $2_{11}-2_{12}$              &   \z4916.319(3)     & \textit{GWi \&\ Churchwell 1975}\\
Formamide     &NH$_2$CHO         & $1_{11}-1_{11}$              &   \z1539.58(2)$^{\rm c}$    & Palmer et al. 1971 \\
              &                  &$2_{11}-2_{12}$               &   \z4618.55(2)$^{\rm c}$    & Rubin et al. 1971\\
Cyanoacetylene&HC$_3$N           & $1-0$                        &    \z9098.6537(3)$^{\rm c}$    & Turner 1971\\ 
              & H$^{13}$C$_3$N   & $1-0$                        &    $\sim9060^{\rm d}$ & \textit{Gardner \&\ GWi 1975a}\\
              & H$^{12+13}$C$_3$N& $1-0$                        &    $\sim9000^{\rm d}$ & \textit{Churchwell et al. 1977}\\
Acetaldehyde  &CH$_3$CHO         &$1_{10}-1_{11}$               &    \z1065.075(5)     & Ball et al. 1971 \\
Methylidine   &CH                &$^2\Pi_{1\over2}, J={1\over2}$  & \z3263.788(10)$^{\rm e}$ & Rydbeck et al. 1973 \\
Thioformaldehyde&H$_2$CS         &$2_{11}-2_{12}$               &   \z3139.38(3)     & Sinclair et al. 1973\\
Methanimine   &CH$_2$NH          &$1_{10}-1_{11}$               &    \z5290.31(4)  & Godfrey et al. 1973 \\
Methylamine   &CH$_2$NH$_2$      &$2_{02}-1_{10}$               &    \z8776.2(1)$^{\rm f}$    & Fourikis et al. 1974\\
Vinyl cyanide &H$_2$CCHCN        &$2_{11}-2_{12}$               &    \z1371.8262(1)     & \textit{Gardner \&\ GWi 1975b}\\
Methyl formate&HCOOCH$_3$-$A$        &$1_{10}-1_{11}$               & \z1610.2445(7)     & Brown et al. 1975 \\
              &HCOOCH$_3$-$E$                  &$1_{10}-1_{11}$    &  \z1610.91(10)     & Churchwell \&\ GWi 1975\\
Dimethyl ether&(CH$_3$)$_2$O      &$2_{02}-1_{11}$               &   \z9119.668(15)$^{\rm c}$    & \textit{GWi \&\ Gardner 1976}\\
Cyanodiacetylene&HC$_5$N         &$4-3$                         &   10650.650(4)     & Avery et al. 1976\\
              &                   &$2-1$                         &  \z5325.328(2)    & \textit{Gardner \&\ GWi 1978}\\
\noalign{\smallskip}\hline
\end{tabular}
}
\end{center}
Only molecules first detected at cm-wavelengths are listed. For early detections in the mm-range, see Snyder 1972.
All molecules were first detected in the transitions listed, except for dimethyl either, which 
was discovered in its $6_{06}-5_{15}$ line at 90938.10(3) MHz by Snyder et al. 1974.
No detections for isotopomers are listed, except for HC$_3$N.\\
Contributions coauthored by GWi are in \textit{italics}.  In most cases these contain essential
laboratory spectroscopy work by GWi.\\
$^{\rm a}$ Apart from the $F=1-1$ line listed, also the $2-2$ line at 1667.3590(2) MHz was reported in the
detection paper. 
$^{\rm b}$ Line with hfs, intensity-weighted frequency is given
$^{\rm c}$ Line with hfs, average of lowest and highest hfs line frequency is given
$^{\rm d}$ Various lines with differently placed $^{13}$C between 8815 and 9100 MHz.
$^{\rm e}$ Apart from the $F=0-1$ line listed, also the $F=1-1$ and $F=1-0$ lines at  
3335.475(10) and 3349.185(10) MHz, resp., were reported in the detection paper (frequencies from Rydbeck et al. 1974). 
$^{\rm f}$ Frequency from Fourikis et al. 1974.
\end{table}

As shown in Table \ref{tab:1}\footnote{The frequencies are taken from the 
JPL catalog (Pickett et al. 1998; see http://spec.jpl.nasa.gov/)  
or the Cologne Database for 
Molecular Spectroscopy 
(M\"uller et al. 2001; see http://www.ph1.uni-koeln.de/vorhersagen/molecules/main\_molecules.html).
Numbers in parentheses are uncertainties in the last significant digit(s).}, 
many of the early molecular detections
were made at low radio frequencies (and toward Sagittarius B2).
A reason for the propensity to find \textit{such}
lines \textit{there} is given in \S \ref{sec:1.1}

Almost concurrently with cm-wavelength molecular 
radio astronomy, shorter wavelength exploration started.
Driven by new high-frequency receiver technology and, generally, 
higher line intensities and a much higher ``line density'' (i.e.
lines per GHz) the ``millimeter explosion'' 
started with the discoveries of CO by R. Wilson et al. and HCO$^+$ 
(``X-ogen'') by Buhl \&\ Snyder, both in 1970 (the latter published in 1971), and continues  through today, having reached the submillimeter range in the 1980s.

Once the millimeter window was opened, (most) people spontaneously lost interest in the cm-range because:
\begin{itemize}
\item Most molecules have the bulk or the entirety of their 
rotational spectrum at (sub)millimeter wavelengths
\item Almost all ``simple'' (di- or triatomic), 
light molecules have (sub)mm or 
far-IR (FIR) rotational spectra
\item Many of the complex molecules were (for a long time) \textit{only} 
found in Sgr B2  
\end{itemize}

Interest remained high in special cm-wavelength lines, including
the NH$_3$ inversion transitions
near 24 GHz, a number of prominent maser lines from OH (hfs, 1.7 GHz and higher
frequencies), H$_2$O (22.2 GHz), CH$_3$OH (25, 12.2, and 6.7 GHz), as well as
the CH hfs lines (3.3 GHz). All of these, except for CH$_3$OH, have all or the 
bulk of their  \textit{rotational} spectrum at submillimeter and far-infrared wavelengths.

\subsection{Sagittarius B2}
\label{sec:1.1}
Considering the mm-wavelength data coming in, 
soon following the radio observations, it became clear that the  emission from many low frequency, low-$J$ 
lines of complex molecules  toward Sgr B2 must be (weakly) inverted. This can be understood 
(at least for linear or near-linear species) from the fact that ``when selection rules for collisional 
transitions in linear molecules are such that dipole collisions do not dominate, the $J=1-0$       
transitions may have population inversion over a wide range of physical conditions 
(see, for example Goldsmith 1972)'' (quoted from Morris et al. 1976). 
This is borne out by their statistical equilibrium  calculations.

However, {\it direct} proof for this assertion came only very recently, from cm-wave 
interferometry of the CH$_3$CHO $1_{11}-1_{10}$ (1065 MHz) and  HC$_3$N $1-0$ (9100 MHz) lines
with the Giant Metre Wave Radio Telescope (GMRT; Chengalur \&\ Kanekar 2003)
and the Australia Telescope Compact Array (ATCA; Hunt et al. 1999), respectively.  The molecular
emission is extended over a 
$2'\times4'$
region and shows a one-to-one correspondence
with the continuum emission, proving that it is inverted. For the CH$_3$CHO 
line a typical
optical depth of 0.035 is derived.

Apart from the mentioned lines, Gardner \&\ Winnewisser (1975b) concluded that 
the vinyl cyanide  line listed in Table \ref{tab:1} is inverted, as did Winnewisser \&\ Gardner (1976) 
for the dimethyl ether line and Sinclair et al. (1973) for the thioformaldehyde line. 
Broten et al. (1976) inferred the same for the 2.663 GHz $J=1-0$ transition of
HC$_5$N.

Once the first complex molecules were found in Sgr B2, this source turned
into a Bonanza! The described inversion mechanism together with a
relatively strong, extended continuum background ``boosted''
many otherwise undetectable lines into ``observability''.
As a consequence, the vast majority of interstellar molecules was and
still is found in 
Sgr B2, many of the most complex ones \textit{only} there. All of the molecules listed in Table \ref{tab:1} 
are found  in this source.

To put GWi's Sgr B2 effort into perspective:
Several of the lines measured by Gardner and him confirmed
``shaky'' detections, based on noisy single line observations.
The vinyl cyanide work presents the first detection of this molecule
in space. The study on different $^{13}$C isotopes  of HC$_3$N 
by Churchwell et al. 
(1977) indicated
chemical fractionation and put a cautionary remark on attempts at 
determining the carbon isotope ratio in the Galactic center.

\subsection{Extended and Compact Organic Molecular Emission in Sgr B2}

Using the (low) rotation temperatures and column densities determined for many of the
species in Table \ref{tab:1}
by Cummins et al. (1986) based on  their 2 -- 3 mm wavelength line survey 
it is completely clear that \textit{all} of the lines in that table must be inverted, which greatly helped their discovery at the time.

Why are these ``old''  Sgr B2 results still highly relevant today?
In large beam observations, such as those of Cummins of al. (angular resolution 
$90''$--$180''$) one observes very low rotation temperatures for many species. 
However, interferometer observations by Snyder and collaborators
with resolutions $< 5''$ indicate $T_{\rm rot} \sim 200$ K for the same species in a very compact region,
dubbed the ``Large Molecule Heimat'' (LMH)  by Snyder et al. (1994).
This source, which is in the immediate vicinity of 
the Sgr B2(N) continuum 
source, is the most prolific source of complex molecules known.

Perhaps most interestingly, in recent years arcsecond resolution
interferometry with the BIMA Array  has resulted in the detection and
imaging of increasingly complex organic species, such as
CH$_2$CHCN, HCOOCH$_3$, and CH$_3$CH$_2$CN; all Miao et al. 1995,
1997), NH$_2$CHO, HNCO, and HCOOCH$_3$ (Kuan \&\ Snyder 1996), CH$_3$COOH
(acetic acid; Mehringer et al. 1997; Remijan et al. 2002), HCOOH
(formic acid; Liu et al. 2001), and (CH$_3$)$_2$CO (acetone; Snyder et
al. 2002); recent single dish detections include CH$_2$OHCHO
(glycolaldehyde, the first sugar; Hollis et al. 2000; 2001);  HOCH$_2$CH$_2$OH 
(ethylene glycol: interstellar ``antifreeze''; Hollis et al. 2002), and 
CH$_2$CHOH (vinyl alcohol: Turner \&\ Apponi 2001).

The elevated complex molecule abundances in a confined, hot, region, such as the LMH, 
may be explainable
by the evaporation of icy grain  mantles, during  which molecules that have grown there to complexity
by continual hydrogenation over a long time are suddenly released into the gas phase
by the ignition of
a high-mass embedded star\footnote{A review of complex molecule formation involving grain mantle processes
has recently been presented by Ehrenfreund \&\ Charnley (2000).}. However, such an event cannot possibly explain
(for luminosity reasons alone) the  unique large-scale distribution of complex molecules in (and maybe around) 
Sgr B2, which is the most prolific source of organic molecules anywhere and, so far,
has  only been  probed by the low frequency lines described above. 
\textit{If} these molecules come off grain mantles, other processes, such as shocks
created, e.g., by cloud-cloud collisions must be invoked. There actually is evidence
for the latter in Sgr B2 (Mehringer \&\ Menten 1997; Sato et al. 2000). 

\subsection{A Giant Organic Cloud around the Galactic Center?}

There is actually  evidence for a giant repository of organic molecules beyond and including 
Sgr B2, coextensive with the Central Molecular Zone (CMZ, see e.g. Morris \&\ Serabyn 1996), 
i.e., stretching from $l = +1\decdeg6$  to $-1\decdeg1$ in a $\sim  \pm  0\deg3$
wide band around the Galactic center. 

The first evidence for extended organic material in the CMZ
came from widespread 4.8 GHz H$_2$CO absorption
(Scoville et al. 1974). Given the ubiquity of formaldehyde 
in molecular clouds, one might dismiss this ``as nothing
special''. Methanol (CH$_3$OH) on the other hand has usually quite
low abundance and is difficult to detect outside hot, dense cloud cores.
Nevertheless, Gottlieb et al. (1979) find the emission in the 834 MHz
$(1^+-1^-)$ line (see Table \ref{tab:1}) in emission and \textit{extended} relative to their $40'$(!)
beam,  concluding it is inverted.

Furthermore, mapping of the HNCO $5_{05}-4_{04}$ transition (made serendipitously
simultaneously with a C$^{18}$O survey), shows that the emission in this
line is extending continuously from $l = -0\decdeg2$ to $+1\decdeg7$ 
(Dahmen et al. 1997). 
The possible existence of such a huge mass of organic material 
in our Galactic center is extremely exciting and its extent, chemistry,
and excitation should be studied with suitable tracers.

\subsection{Randbemerkung: TMC-1}
After Sgr B2, GWi and his collaborators turned their attention to complex molecules in cold dark clouds.
In the abstract of a paper by Churchwell, Winnewisser, \&\ Walmsley (1978) 
titled 
fetchingly ``Molecular observations of a possible proto-solar nebula in a dark cloud in Taurus'', 
these authors
reported ``a small molecular condensation near the south eastern edge of Heiles' cloud 2''
in which they find ``strong emission from the $J=9-8$ transition of HC$_5$N and the
$J=1-0$ transition of HC$_3$N'', and go on to ``refer to this small cloud as
the Taurus Molecular Cloud 1, or TMC 1''.
It turned out that they started a vogue:
As of this writing (2003 October 25) the Astronomical Data 
System\footnote{http://adsabs.harvard.edu/} lists a total
of 267 abstracts containing TMC 1 in their text. 

Churchwell et al. were by no means the first to notice this spectacular source, which lies
in the heart of the Taurus molecular cloud complex, 
and is well known for its high abundance of linear carbon chains, in particular
the cyanopolyynes, the longest 
of which, HC$_{11}$N has \textit{only} been found there (Bell et al. 1997). Citing them: ``Morris
\footnote{Mark Morris tells me that the first person to point out TMC 1 to him was
Nick Scoville.} et al. (1976)
first drew attention to this region by their detection of the $J=5-4$ transition of
HC$_3$N; this was followed by the detection of the $J=4-3$ transition of HC$_5$N
by MacLeod et al. (1978)[sic]\footnote{That paper actually appeared in 1979.} and the $J=9-8$ transition of HC$_7$N by Kroto et al. (1977)''.

A great amount of work on TMC 1 was done by the group at the Herzberg Institute of Astrophysics, GWi's onetime home,
and at the Nobeyama Radio Observatory (NRO), whose 45 m telescope is ideally suited for the
22 -- 50 GHz range in which many heavy species at  TMC 1's cold (10 K) temperature have their 
strongest emission. In a single paper, Kaifu et al. (1987) reported seven(!), strong(!)
unidentified lines, which, aided by laboratory work of  Saito et al. (1987) and Yamamoto et al. (1987),
led to the first identifications of  CCS and CCCS.

The NRO's TMC 1 work culminated with a complete 8800 -- 50000 MHz line survey,
described by Ohishi \&\ Kaifu (1998), but unfortunately not (yet) published in its entirety,
which contains a total of 404 lines from 38 molecules (not counting isotopomers). Eleven of the molecules were newly identified in the course of the survey. 
It is a great testimony to the 
laboratory spectroscopists involved that only a single line remains unidentified!

\section{Light Intermezzo -- Interstellar Hydrides}
When I was a student, an article by G.  Winnewisser, Aliev \&\ Yamada (1982)
drew my attention. To quote from its abstract:
``The great progress molecular astronomy has made in the past decade 
has focused attention to one special class of molecules: 
the hydrides. The presently known spectra of the hydrides are reviewed, 
both interstellar and laboratory. No interstellar metal hydrides 
are presently known. A large number of important hydrides 
exhibit spectra only in the submillimetre and far infrared region. 
Further technical advances in this area are expected to yield 
new interstellar detections.'' 

\subsection{Metal-bearing Molecules in the Interstellar Medium}
Observations of species containing metals, as well as other
refractory species such as SiO, deliver important information
on depletion into dust grains. Moreover, when released  
either from sputtering of the grains or their mantles by
violent dynamics, such as protostellar outflows, they can be used
as probes of these phenomena (see, e.g. Bachiller 1996). 
 
Today the third sentence quoted above, ``No 
interstellar metal hydrides $\ldots$''
is, unfortunately, still true, if we use the strict (chemistry) meaning of 
``metal'' (e.g. Na, K, Mg, Al, Fe,$\ldots$). Four 
diatomic, non H-bearing metal 
compounds (NaCl, KCl, AlCl, and AlF and three triatomic
ones  (MgNC, MGCN, NaCN, and AlNC)   have been detected  
in the dense envelope of the
carbon star IRC+10216 (Cernicharo \&\ Gu\'elin 1987; Gu\'elin et al. 1993; Kawaguchi et al. 1993; 
Ziurys et al. 1995; Turner et al. 1994; Gu\'elin et al. 1996; Ziurys et al. 2002). 
MgNC, NaCN, AlF, and NaCl were also found, recently, 
toward the proto-planetary nebula (PN) 
CRL 2688 (Highberger et al. 2001, 2003; Ziurys et al. 2002; see Petrie 1999 for a discussion of other 
candidate species for detection). Still, only one metal-containing molecule, FeO, has been 
detected in the interstellar medium, toward Sgr B2 (Walmsley et al. 2002).

As to metal hydrides: Radio and (sub)millimeter wavelength searches 
for interstellar NaH (Plambeck \&\ Erickson 1982) and  LiH\footnote{The ground state $J=1-0$ transition
of LiH
is near 444 GHz, an atmospherically very  unfavorable 
frequency. A tentative detection of this line
was made in the redshifted ($z=0.68$) dense absorbing cloud
toward the B0218+357 gravitational lens system (Combes \&\ Wiklind  1998).}
have so far been unsuccessful,
as was a search for  MgH in IRC+10216's envelope 
(Avery et al. 1994). Optical wavelength absorption from
CaH, AlH, FeH, and MgH has been found in cool stars and the Sun,
MgH even in the integrated light of external galaxies (see, e.g., Spinrad
\& Wood 1965)

\subsection{Light Hydrides -- Present Status}
The history of interstellar hydrides begins with the detection 
by Dunham (1937) of
three optical absorption lines  at 3957.7, 4232.6, and 4300.3 \AA\
from diffuse interstellar clouds, which subsequently were found
to arise from CH (4300.3 \AA; McKellar 1940; suggested by Swings \&\
Rosenfeld 1937) and CH$^+$ (the former two; Douglas \&\ Herzberg 1941).

It has been thought for a long time that hydrides might play an important r\^ole
in cooling the highest density regions of interstellar clouds. 
Hydrides have, firstly, large level spacings,
allowing them to emit a relatively large amount of energy
per quantum. Secondly, their large dipole moments require
high critical densities for their excitation. Thus, their rotational levels
are only excited in high density regions with $n > 10^6$ cm$^{-3}$, making 
them effective coolants in this regime, where even rare CO isotopes
are so optically thick that CO cooling becomes negligible compared
to that of (the hydride) H$_2$O. 

This regime
was excluded in the classical study by Goldsmith and Langer (1978)
on ``Molecular Cooling and Thermal Balance in Dense Molecular Clouds'',
but was re-addressed by Neufeld et al. (1995), who found that for
elevated  temperatures, $T = 100$ K, H$_2$O cooling dominates and
the hydrides' contribution is comparable to that \textit{of all other
molecules combined}, i.e. $\approx 30$ \%\  of the former. 

Recent detections of hydride rotational transitions were made with 
the Caltech Submillimeter Observatory (CSO), namely H$_3$O$^+$ [$J_K=1^-_1-2^+_1,3^+_2-2^-_2,3^+_0-2^-_0$/307, 365, 396 GHz, Wootten et al. 1991;
Phillips et al. 1992], 
SiH ($^2\Pi_{1\over2}, J={3\over2}-{1\over2}$/625, 628 GHz; Schilke et al. 2001)\footnote{Tentative detection}, 
HCl ($J=1-0$/626 GHz); Blake et al. 1985),
H$_2$O, H$_2^{18}$O ($1_{10}-1_{01}$/557, 548 GHz; Melnick et al. 2000),
HDO (the $1_{01}-0_{00}$/465 GHz  and $1_{11}-0_{00}$/894 GHz ground-state transitions; Schulz et al.
1991, Pardo et al. 2001)\footnote{A number of higher excitation cm and mm-wavelength H$_2$O, H$_2^{18}$O, 
  and HDO transitions have also been found.}, and NH$_2$ 
($J={3\over2}-{3\over2},{1\over2}-{1\over2},
{3\over2}-{1\over2}$/462, 459, 461 GHz); van Dishoeck et al. 1993).
H$_2$D$^+$ ($1_{10}-1_{11}$/372 GHz), the 
deuterated isotope of the key H$_3^+$ molecule was discovered with the James-Clerck-Maxwell telescope 
(Stark et al. 1999)\footnote{H$_3^+$ itself was finally detected, 
after many arduous  attempts, in absorption 
at near-infrared
wavelengths by Geballe \&\ Oka (1996), benefitting from  a monumental body
of laboratory work, mainly gathered by Oka and collaborators 
(see these proceedings).}.
Observations with the Long Wavelength Spectrometer aboard 
the Infrared Space Observatory led to the discoveries of
HF ($J=2-1$/2.47 THz; Neufeld et al. 1997), HD ($J=1-0$/2.68 THz and $2-1$/5.36 THz;
Polehampton et al. 2002), plus various H$_2$O, OH, and CH 
rotational and some H$_3$O$^+$ inversion transitions
(many toward Sgr B2; Goicoechea \& Cernicharo 2001 and these proceedings).

A preliminary conclusion is that (for the only two sources observed, Orion-KL and Sgr B2) the
measured HCl and HF abundances indicate that only a few percent of all chlorine 
is in these molecules, suggesting significant depletion. If this should hold true for
other hydrides as well, their cooling contribution could be much smaller than 
discussed above.  More sensitive observations of more transitions toward more lines 
of sight are needed. These will be possible with the Herschel Space 
Observatory\footnote{http://astro.estec.esa.nl/SA-general/Projects/First/first.html} and the 
Stratospheric Observatory For Infrared Astronomy (SOFIA\footnote{http://sofia.arc.nasa.gov/}).

Under GWi's direction the KOSMA  Spectroscopy Group has 
produced definitive submillimeter spectroscopy of NH$_2$, C, $^{13}$C, NH, 
SH, SD, and CF.

\section{The quest for high spectroscopic precision --
``Deutsche Gr\"undlichkeit'' gone loose or urgent necessity?}
\label{sec:2}

Over the years, the KOSMA lab, with GWi as the driving force, has produced 
``definitive'' spectroscopy (i.e. frequencies to a few times 10 kHz rms 
up to 1 THz or higher) for a large number of astrophysically important 
molecules, including the first useful data on many \textit{rare isotopomers} 
and various \textit{vibrationally excited states}.              

These  are\footnote{For references, see 
http://www.ph1.uni-koeln.de/vorhersagen/catalog/catdir.html}: 
HCS$^+$, HC$^{14+15}$N, $^{12+13}$CO, C$^{17+18}$O, H$_2^{12+13}$C$^{16+18}$O, 
ground-state + vibrationally excited 
H$^{12+13}$C$_3^{14+15}$N,
C$_2$H$_5$OCH$3$, CCS,  CCC,  ground-state + vibrationally excited  HNC, 
CH$2$NH, CH$_3$CCH, CCH, $^{32+33}$SO$_2$, S$^{16+17}$O$_2$, 
C$^{14+15}$N, CH$_2$CCHCN, CCCO, CH$_3$C$_4$H, NH$_2$D (hfs), HC$_5$N, HNCO,
and the seven simple species with submillimeter spectra listed above\footnote{Note the tenacity with which cyanoacetylene
was pursued!; cf. \S \ref{sec:1}}. 

The first astronomical detections of C$^{13}$CH and 
HC$^{15}$N were made by  the Cologne group 
with the KOSMA telescope (Saleck et al. 1992, 1993) and of $^{13}$C$^{17}$O
with the Swedish-ESO-Submillimeter Telescope (Bensch et al. 2001).

Among others, the Cologne spectroscopy effort is characterized by,
first, comprehensive  presentations of the \textit{complete}
spectrum of a given species, in general up to higher 
(several THz) frequencies than many other listings, and, second,
significantly higher accuracy data for many ``common'' molecules,
such as CO and its isotopomers, than available before.

Here we argue, using two examples, that both high frequency \textit{accuracy}
and \textit{comprehensiveness} are necessary for progress in various
different areas.

\subsection{Accuracy}
Ambipolar diffusion -- ion/neutral slip -- is an important mechanism
in the star-formation process by which magnetic fields initially supporting clouds
against collapse slip relative to the neutral gas, leading to the formation of dense cloud cores
(see, e.g. Shu et al. 1987).
Theory predicts velocity differences $<1$ km~s$^{-1}$, 
corresponding to  1 MHz at 300 GHz,    
between ionic and neutral species (Roberge et al. 1995). 
In particular, at 
submillimeter wavelengths 
present  frequency uncertainties for 
many potentially suitable lines are much too high to draw any reliable conclusions.

\subsection{Comprehensiveness and Line Confusion}
In a recent paper, Kuan et al. (2003) claim to have detected the simplest
amino acid, glycine (H$_2$NCH$_2$COOH) in three warm, dense interstellar clouds,
including Sgr B2. If true, this would be the conclusion of a long quest
(see, e.g., Snyder et al. 1983; Berulis, Winnewisser, Krasnov, \&\ 
Sorochenko 1985; Combes  et al. 1996; Ceccarelli et al. 2000). 

Any such ``identification'' must rely on a large number of lines.
Even so (or maybe \textit{because of this}) the \textit{confusion problem} is a formidable one.
Inspecting any figure presented by Kuan et al. we find comparably many
glycine as unidentified (``U'') lines. 
For a wider view, see the 218 -- 263 GHz line survey by Nummelin et al.
(1998, 2000), which, for Sgr B2(N), is at or near the confusion limit throughout that band, 
even at their relatively moderate sensitivity (compared
to the interferometric data by Snyder and his collaborators).

In the high density ($n\sim 10^7$ cm$^{-3}$) Sgr B2 ``Large Molecule Heimat'' (see \S \ref{sec:1.1}) 
level populations
are in Local Thermodynamic Equilibrium characterized by 
rotation temperatures, $T_{\rm rot}$, around 200 K (Liu et al. 2001).
At this temperature, the lower vibrationally excited states
of a large number of complex molecules are populated, giving
rise to a plethora of emission lines, probably with comparable
intensities to the putative glycine lines. The width of the LMH's lines,
typically $\sim6$ -- $10$ km~s$^{-1}$, adds additional complication. 
Given the gigantic partition functions, many of the lines are
weak.

In my opinion, it  is completely impossible to conclusively 
identify \textit{very} complex molecules in a ``line jungle'', 
such the one found in the Sgr B2 Large Molecule Heimat\footnote{High spatial resolution interferometry 
as employed by L. Snyder and collaborators helps significantly
in situations as complicated as the LMH:
Individual molecules or groups of molecules in general have slightly 
different spatial distributions, so that imaging adds an additional
dimension to a confident identification.}, 
if one has not a much better knowledge of, a.o., vibrationally 
excited lines of many more species than is available today.
This means that much more (tedious and little-appreciated) lab work is needed
to make progress here: A big job for the Cologne group, which has already
measured the spectra (although not necessarily those of the 
vibrationally excited states) of a number of extant molecules.

\subsection*{Apology}
This contribution concerned itself mostly with (some aspects of) 
GWi's early astronomical career. Among many others, 
I didn't discuss such interesting and important topics as
\begin{itemize}
\item NH$_3$ hfs anomalies (Stutzki et al. 1982, 1984; Stutzki \&\ Winnewisser 1985a,b,c)
 
\item his pushing the Cologne Acousto-Optical Spectrometers and receivers
to become  world-class, frontline instrumentation (see, e.g., 
Schieder et al. 1989;
Klumb et al. 1995; Graf et al. 1998, 2003; see also 
Gary Melnick's SWAS contribution in these proceedings)

\item any of all the nice KOSMA/Gornergrat results

\item any of his non-astronomical spectroscopy; little pre-Cologne spectroscopy.
\end{itemize}

\noindent
\textit{Acknowledgement:} I would like to thank Malcolm Walmsley and Lew Snyder
for their encouragement.
%
%
%
%
%



\printindex
\end{document}